# 後量子密碼學神經網路
## Post-Quantum Cryptography Neural Network

**Abel C. H. Chen**

**Telecommunication Laboratories, Chunghwa Telecom Co., Ltd.**

## 摘要

近年來,量子電腦和 Shor 量子演算法已經對現行主流的非對稱式密碼學(包含 RSA 和橢圓曲線密碼學(Elliptic Curve Cryptography, ECC))造成威脅。因此,需要建構一套後量子密碼學(Post-Quantum Cryptography, PQC)方法來抵抗量子計算攻擊。有鑑於此,本研究提出一套基於編碼的後量子密碼學神經網路,可把後量子密碼學方法對應到神經網路結構,並且搭配非線性激活函數、隨機數擾動密文、以及約束密文呈均勻分佈提升密文的安全性。在實驗中,本研究以行動網路訊號為例進行實證,可以通過本研究提出的後量子密碼學神經網路進行加密和解密,並且可以保證密文服從均勻分佈。未來可將後量子密碼學神經網路推廣到各式各樣的應用。

**關鍵詞:後量子密碼學、McEliece 密碼學、神經網路**

*通訊作者: Abel C. H. Chen (chchen.scholar@gmail.com)

## Abstract

In recent years, quantum computers and Shor's quantum algorithm have posed a threat to current mainstream asymmetric cryptography methods (e.g. RSA and Elliptic Curve Cryptography (ECC)). Therefore, it is necessary to construct a Post-Quantum Cryptography (PQC) method to resist quantum computing attacks. Therefore, this study proposes a PQC-based neural network that maps a code-based PQC method to a neural network structure and enhances the security of ciphertexts with non-linear activation functions, random perturbation of ciphertexts, and uniform distribution of ciphertexts. In practical experiments, this study uses cellular network signals as a case study to demonstrate that encryption and decryption can be performed by the proposed PQC-based neural network with the uniform distribution of ciphertexts. In the future, the proposed PQC-based neural network could be applied to various applications.

# 壹、前言

近年來，隨著量子電腦的發展和Shor量子演算法(Shor, 1997)的提出，已經可以結合量子特性在多項式時間解決部分的NP (Non-deterministic Polynomial-time)問題，如：質因數分解問題、離散對數問題等。因此，目前主流非對稱式密碼學將面臨量子計算攻擊，包含以質因數分解問題為基礎的RSA密碼學和以離散對數問題為基礎的橢圓曲線密碼學(Elliptic Curve Cryptography, ECC)。有鑑於此，美國國家標準技術局(National Institute of Standards and Technology, NIST)開始公開徵求後量子密碼學 (Post-Quantum Cryptography, PQC)方法(Alagic et al., 2022)，並整理了基於格的密碼學(Nejatollahi et al., 2019)、基於編碼的密碼學(McEliece, 1978; Thiers & Freudenberger, 2022)、基於多變量的密碼學(Alagic et al., 2022)、基於雜湊的密碼學(Sahraneshin et a., 2023)等，分別可以在不同的NP問題的基礎上提供安全性。例如，基於編碼的密碼學中的McEliece密碼學主要建構多個矩陣作為私鑰，把矩陣相乘後的矩陣作為公鑰，建構在非負矩陣分解(Non-negative Matrix Factorization, NMF)這個NP問題的基礎上，並且可以對密文加入雜訊，再通過糾錯碼去除雜訊，提升抵抗量子計算攻擊的能力。

除此之外，神經網路在近年來蓬勃發展，並且具有良好的編碼和解碼能力。有鑑於此，本研究提出後量子密碼學神經網路，結合神經網路編解碼的能力(Guo et al., 2023)，並且加入雜訊產生密文。本研究將先借鑑基於編碼的後量子密碼學方法，建構符合後量子密碼學方法的神經網路結構，再搭配非線性激活函數、隨機數擾動密文、以及約束密文呈均勻分佈等面向來打造後量子密碼學神經網路。

本研究的主要創新點條列如下：

1. 提出後量子密碼學神經網路，建構符合基於編碼後量子密碼學的神經網路結構，可以將輸入層到密文層間的權重集合作為公鑰加密使用，並且將密文層到輸出層之間的權重集合作為私鑰加密使用。
2. 提出增加密文混亂程度的策略(1)，本研究在密文層加入隨機數擾動，並且可以在解密階段去除隨機數影響，還原成原本的明文。
3. 提出增加密文混亂程度的策略(2)，本研究在讓密文層的輸出值盡可能服從均勻分佈，在損失函數加入卡方分佈的累加分配函數(Cumulative Distribution Function, CDF)值約束神經網路產出足夠均勻的密文層值。

本文總共分為五個章節。第貳節介紹基於編碼後量子密碼學，並以 McEliece 密碼學為例說明其原理與計算例。第參節詳述本研究提出的後量子密碼學神經網路，從對應 McEliece 密碼學的神經網路結構到改進版的神經網路結構，並且設計密文層隨機數擾動和結合檢定均勻分佈的損失函數，提升密文的安全性。第肆節以真實資料進行證分析，驗證本研究提出的後量子密碼學神經網路可以用來加密行動網路訊號，並且討論在不同參數值時對明文誤差和密文均勻分佈的影響。最後，第伍節總結本研究的貢獻，並討論未來的研究方向。

# 貳、基於編碼後量子密碼學

本節介紹基於編碼後量子密碼學的 McEliece 密碼學(McEliece, 1978)，將以漢明

(Hamming)糾錯碼為例作為 McEliece 密碼學的編碼方法，再說明 McEliece 密碼學加解密的作法及其原理。

**一、漢明糾錯碼方法**

本節將分別從初始階段、編解碼、以及糾錯三個章節來介紹漢明糾錯碼方法(McEliece, 1978)。

**(一) 初始階段**

在初始階段，漢明糾錯碼方法將先產生三個矩陣分別是：糾錯碼生成矩陣(generator matrix) $G$、對應糾錯碼生成矩陣 $G$ 的糾錯矩陣 $H$、以及對應糾錯碼生成矩陣 $G$ 的解碼矩陣 $R$。為說明漢明糾錯碼方法，本文提供計算例，生成矩陣 $G$、糾錯矩陣 $H$、解碼矩陣 $R$ 分別如公式(1)、(2)、(3)所示。

$$G = \begin{bmatrix} 1 & 1 & 1 & 0 & 0 & 0 & 0 \\ 1 & 0 & 0 & 1 & 1 & 0 & 0 \\ 0 & 1 & 0 & 1 & 0 & 1 & 0 \\ 1 & 1 & 0 & 1 & 0 & 0 & 1 \end{bmatrix} \quad (1)$$

$$H = \begin{bmatrix} 1 & 0 & 0 \\ 0 & 1 & 0 \\ 1 & 1 & 0 \\ 0 & 0 & 1 \\ 1 & 0 & 1 \\ 0 & 1 & 1 \\ 1 & 1 & 1 \end{bmatrix} \quad (2)$$

$$R = \begin{bmatrix} 0 & 0 & 0 & 0 \\ 0 & 0 & 0 & 0 \\ 1 & 0 & 0 & 0 \\ 0 & 0 & 0 & 0 \\ 0 & 1 & 0 & 0 \\ 0 & 0 & 1 & 0 \\ 0 & 0 & 0 & 1 \end{bmatrix} \quad (3)$$

**(二) 編解碼**

在編碼的過程中，主要將原始資料 $x$ 與生成矩陣 $G$ 相乘，取得編碼後資料 $y$，如公式(4)所示。並且，在解碼的過程中，可以將編碼後資料 $y$ 與解碼矩陣 $R$ 相乘，取得原始資料 $x$，如公式(5)所示。本文以 $x = [1 \ 0 \ 0 \ 0]$ 作為計算例進行演算說明，其編碼後資料 $y$ 為 $[1 \ 1 \ 1 \ 0 \ 0 \ 0 \ 0]$，如公式(6)所示，並可通過編碼矩陣 $R$ 解碼回 $x$，如公式(7)所示。

$$y = xG \quad (4)$$
$$x = yR \quad (5)$$
$$y = xG = [1 \ 0 \ 0 \ 0] \begin{bmatrix} 1 & 1 & 1 & 0 & 0 & 0 & 0 \\ 1 & 0 & 0 & 1 & 1 & 0 & 0 \\ 0 & 1 & 0 & 1 & 0 & 1 & 0 \\ 1 & 1 & 0 & 1 & 0 & 0 & 1 \end{bmatrix} = [1 \ 1 \ 1 \ 0 \ 0 \ 0 \ 0] \quad (6)$$

$$x = yR = [1\ 1\ 1\ 0\ 0\ 0\ 0]\begin{bmatrix}0 & 0 & 0 & 0\\0 & 0 & 0 & 0\\1 & 0 & 0 & 0\\0 & 0 & 0 & 0\\0 & 1 & 0 & 0\\0 & 0 & 1 & 0\\0 & 0 & 0 & 1\end{bmatrix} = [1\ 0\ 0\ 0] \quad (7)$$

**(三) 糾錯**

為演示糾錯方法，本文加入隨機數 $r$ 到編碼後資料 $y$ 中，並得到帶有錯誤訊息(或稱為雜訊)的 $y'$，如公式(8)所示。帶有雜訊的 $y'$ 可以與糾錯矩陣 $H$ 相乘得到錯訊訊息的位置 $z$ (如公式(9)所示)，並且進行更正。本文以 $r = [0\ 0\ 0\ 0\ 0\ 0\ 1]$ 作為雜訊來計算帶有雜訊的 $y'$，如公式(10)所示。通過與糾錯矩陣 $H$ 相乘得到 $z = [1\ 1\ 1]$ (如公式(11)所示)，表示第 7 個 bit 有錯，更正 $y'$ 第 7 個 bit 後得到 $y = [1\ 1\ 1\ 0\ 0\ 0\ 0]$。

$$y' = y + r \quad (8)$$
$$z = y'H \quad (9)$$
$$\begin{aligned}y' = y + r &= [1\ 1\ 1\ 0\ 0\ 0\ 0] + [0\ 0\ 0\ 0\ 0\ 0\ 1]\\ &= [1\ 1\ 1\ 0\ 0\ 0\ 1]\end{aligned} \quad (10)$$

$$z = y'H = [1\ 1\ 1\ 0\ 0\ 0\ 1]\begin{bmatrix}1 & 0 & 0\\0 & 1 & 0\\1 & 1 & 0\\0 & 0 & 1\\1 & 0 & 1\\0 & 1 & 1\\1 & 1 & 1\end{bmatrix} = [1\ 1\ 1] \quad (11)$$

**二、McEliece 密碼學方法**

本節將分別從金鑰產製、加密、以及解密三個章節來介紹 McEliece 密碼學方法。

**(一) 金鑰產製**

在金鑰產製過程中，McEliece 密碼學方法將產生三個矩陣分別是：擾碼器(scrambler)矩陣 $S$、糾錯碼生成矩陣 $G$、以及置換矩陣(permutation matrix) $P$，將這三個矩陣作為私鑰(private key)。並且，將三個矩陣相乘得到公鑰 $G'$，如公式(12)所示。為說明 McEliece 密碼學方法，本文提供計算例，擾碼器矩陣 $S$、生成矩陣 $G$、置換矩陣 $P$ 分別如公式(13)、(1)、(14)所示，其中生成矩陣 $G$ 採用第貳.一.(一)節的例子。通過公式(15)可以取得公鑰 $G'$。

$$G' = SGP \quad (12)$$

$$S = \begin{bmatrix}1 & 1 & 0 & 1\\1 & 0 & 0 & 1\\0 & 1 & 1 & 1\\1 & 1 & 0 & 0\end{bmatrix} \quad (13)$$

$$P = \begin{bmatrix}0 & 1 & 0 & 0 & 0 & 0 & 0\\0 & 0 & 0 & 1 & 0 & 0 & 0\\0 & 0 & 0 & 0 & 0 & 0 & 1\\1 & 0 & 0 & 0 & 0 & 0 & 0\\0 & 0 & 1 & 0 & 0 & 0 & 0\\0 & 0 & 0 & 0 & 0 & 1 & 0\\0 & 0 & 0 & 0 & 1 & 0 & 0\end{bmatrix} \quad (14)$$

$G' = SGP$

$$= \begin{bmatrix} 1 & 1 & 0 & 1 \\ 1 & 0 & 0 & 1 \\ 0 & 1 & 1 & 1 \\ 1 & 1 & 0 & 0 \end{bmatrix} \begin{bmatrix} 1 & 1 & 1 & 0 & 0 & 0 & 0 \\ 1 & 0 & 0 & 1 & 1 & 0 & 0 \\ 0 & 1 & 0 & 1 & 0 & 1 & 0 \\ 1 & 1 & 0 & 1 & 0 & 0 & 1 \end{bmatrix} \begin{bmatrix} 0 & 1 & 0 & 0 & 0 & 0 & 0 \\ 0 & 0 & 0 & 1 & 0 & 0 & 0 \\ 0 & 0 & 0 & 0 & 0 & 0 & 1 \\ 1 & 0 & 0 & 0 & 0 & 0 & 0 \\ 0 & 0 & 1 & 0 & 0 & 0 & 0 \\ 0 & 0 & 0 & 0 & 0 & 1 & 0 \\ 0 & 0 & 0 & 0 & 1 & 0 & 0 \end{bmatrix} \quad (15)$$

$$= \begin{bmatrix} 0 & 1 & 1 & 0 & 1 & 0 & 1 \\ 1 & 0 & 0 & 0 & 1 & 0 & 1 \\ 1 & 0 & 1 & 0 & 1 & 1 & 0 \\ 1 & 0 & 1 & 1 & 0 & 0 & 1 \end{bmatrix}$$

**(二) 加密**

在加密的過程中，主要將明文 $x$ 與公鑰 $G'$ 相乘，並且為了讓相同明文可以對應到不同的密文，再加入一個隨機數 $r$ 進行擾動，計算得到密文 $y$，如公式(16)所示。本文以 $x = [1 \ 0 \ 0 \ 0]$ 和 $r = [0 \ 0 \ 0 \ 0 \ 0 \ 0 \ 0]$ 作為計算例進行演算說明，其計算後密文 $y$ 為 $[0 \ 1 \ 1 \ 0 \ 1 \ 0 \ 1]$，如公式(17)所示。其中，為簡化說明，本例令隨機數 $r$ 的元素為 0，當如果隨機數 $r$ 的元素為 1 時,可通過第貳.一.(三)節糾錯去除隨機數 $r$ 的影響。

$$y = xG' + r = xSGP + r \quad (16)$$

$$y = xG' + r$$
$$= [1 \ 0 \ 0 \ 0] \begin{bmatrix} 0 & 1 & 1 & 0 & 1 & 0 & 1 \\ 1 & 0 & 0 & 0 & 1 & 0 & 1 \\ 1 & 0 & 1 & 0 & 1 & 1 & 0 \\ 1 & 0 & 1 & 1 & 0 & 0 & 1 \end{bmatrix} + [0 \ 0 \ 0 \ 0 \ 0 \ 0 \ 0] \quad (17)$$
$$= [0 \ 1 \ 1 \ 0 \ 1 \ 0 \ 1]$$

**(三) 解密**

由於採用公鑰 $G'$ 加密得到密文，而 $G'$ 為 $S$、$G$、$P$ 三個矩陣相乘的矩陣，所以解密的過程主要乘上對應的反矩陣或解碼矩陣還原明文。首先，把密文乘上置換矩陣 $P$ 的反矩陣，去除密文中的 $P$ 矩陣，如公式(18)所示。通過第貳.一.(三)節糾錯方法，可以去除隨機數 $r$ 的影響(即消除 $rP^{-1}$)，取得 $xSG$。再通過第貳.一.(二)節解碼方法，乘上編碼矩陣 $R$，去除密文中的 $G$ 矩陣，如公式(19)所示。最後把密文乘上擾碼器矩陣 $S$ 的反矩陣，去除密文中的 $S$ 矩陣，即可完成解密，如公式(20)所示。

$$yP^{-1} = xG'P^{-1} + rP^{-1}$$
$$= [0 \ 1 \ 1 \ 0 \ 1 \ 0 \ 1] \begin{bmatrix} 0 & 0 & 0 & 1 & 0 & 0 & 0 \\ 1 & 0 & 0 & 0 & 0 & 0 & 0 \\ 0 & 0 & 0 & 0 & 1 & 0 & 0 \\ 0 & 1 & 0 & 0 & 0 & 0 & 0 \\ 0 & 0 & 0 & 0 & 0 & 0 & 1 \\ 0 & 0 & 0 & 0 & 0 & 1 & 0 \\ 0 & 0 & 1 & 0 & 0 & 0 & 0 \end{bmatrix} \quad (18)$$
$$= [1 \ 0 \ 1 \ 0 \ 1 \ 0 \ 1] = xSGPP^{-1} + rP^{-1} = xSG + rP^{-1}$$

$$xSGR = [1 \ 0 \ 1 \ 0 \ 1 \ 0 \ 1] \begin{bmatrix} 0 & 0 & 0 & 0 \\ 0 & 0 & 0 & 0 \\ 1 & 0 & 0 & 0 \\ 0 & 0 & 0 & 0 \\ 0 & 1 & 0 & 0 \\ 0 & 0 & 1 & 0 \\ 0 & 0 & 0 & 1 \end{bmatrix} \quad (19)$$

$$\begin{aligned}
&= \begin{bmatrix} 1 & 1 & 0 & 1 \end{bmatrix} = xS \\
xSS^{-1} &= \begin{bmatrix} 1 & 1 & 0 & 1 \end{bmatrix} \begin{bmatrix} 1 & 1 & 0 & 1 \\ 1 & 1 & 0 & 0 \\ 0 & 1 & 1 & 1 \\ 1 & 0 & 0 & 1 \end{bmatrix} \\
&= \begin{bmatrix} 1 & 0 & 0 & 0 \end{bmatrix} = x
\end{aligned} \quad (20)$$

McEliece 密碼學方法中的編解碼方法除了採用漢明糾錯碼方法外，也可以採用其他非線性的糾錯碼方法來提升安全性。並且，本文為提供計算例說明，所以僅採用小矩陣計算，在實務上可以採用大矩陣，可以加密較多的資訊，並且也可以提升安全性。例如，目前在美國國家標準技術局評選後量子密碼學標準的 McEliece 密碼學方法採用不同大小的大矩陣來達到標準要求的安全等級。

## 參、後量子密碼學神經網路

為詳細說明本研究提出的後量子密碼學神經網路，本文以循序漸進的方式先介紹對應漢明糾錯碼方法的神經網路，再介紹對應 McEliece 密碼學方法的神經網路，最後再說明本研究提出的後量子密碼學神經網路。

### 一、對應漢明糾錯碼方法的神經網路

本節說明對應漢明糾錯碼方法的神經網路，以第貳.一節中的例子對應神經網路結構，如圖 1 所示。輸入層為 4 個神經元、隱藏層為 7 個神經元、以及輸出層為 4 個神經元；其中，每個神經元不考慮偏差項(bias)，激活函數採用線性函數，並且輸出層的值與輸入層的值一樣(即自編碼器)。因此，輸入層和隱藏層之間的權重矩陣為 4 x 7 大小的矩陣 $G$ (對應生成矩陣 $G$)，權重值如公式(1)所示；隱藏層和輸出層之間的權重矩陣為 7 x 4 大小的矩陣 $R$ (對應解碼矩陣 $R$)，權重值如公式(3)所示。在本節中，神經元之間的權重為 1 時採用紅線表示，神經元之間的權重為 0 時採用黑線表示，即可把第貳.一節中的漢明糾錯碼方法對應到神經網路(即自編碼器)，可進行編碼和解碼。當輸入層的值為 $x = \begin{bmatrix} 1 & 0 & 0 & 0 \end{bmatrix}$，隱藏層 $y$ 的值為 $\begin{bmatrix} 1 & 1 & 1 & 0 & 0 & 0 & 0 \end{bmatrix}$ (即編碼後的值)，並且可以通過輸出層還原到原本的值 $\begin{bmatrix} 1 & 0 & 0 & 0 \end{bmatrix}$ (即解碼後的值)。

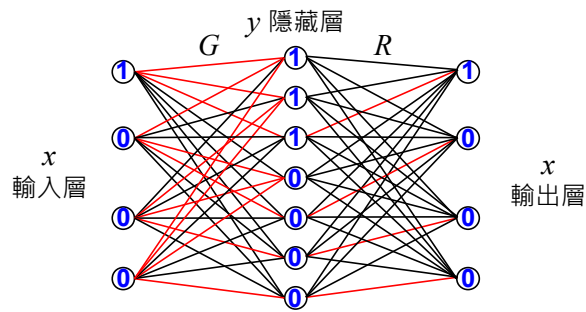

圖 1、對應漢明糾錯碼方法的神經網路

### 二、對應 McEliece 密碼學方法的神經網路

本節說明對應 McEliece 密碼學方法的神經網路，以第貳.二節中的例子對應神經網

路結構，如圖 2 所示。輸入層為 4 個神經元作為明文、第 3 層隱藏層為 7 個神經元作為密文、以及輸出層為 4 個神經元作為明文；其中，每個神經元不考慮偏差項(bias)，激活函數採用線性函數，並且輸出層的值與輸入層的值一樣(即自編碼器)。為簡化表示方式，神經元之間的權重為 1 時採用紅線表示，神經元之間的權重為 0 時採用黑線表示。

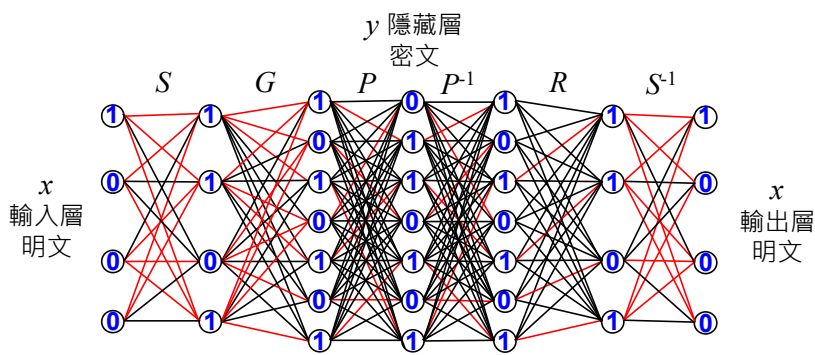

圖 2、對應 McEliece 密碼學方法的神經網路

　　加密的計算為輸入層到第 3 層隱藏層。其中，輸入層和第 1 層隱藏層之間的權重矩陣為 4 x 4 大小的矩陣 $S$ (對應擾碼器矩陣 $S$)，權重值如公式(13)所示；第 1 層隱藏層和第 2 層隱藏層之間的權重矩陣為 4 x 7 大小的矩陣 $G$ (對應生成矩陣 $G$)，權重值如公式(1)所示；第 2 層隱藏層和第 3 層隱藏層之間的權重矩陣為 7 x 7 大小的矩陣 $P$ (對應置換矩陣 $P$)，權重值如公式(14)所示。當輸入層的值為$x = [1\ \ 0\ \ 0\ \ 0]$，第 3 層隱藏層 $y$ 的值為$[0\ \ 1\ \ 1\ \ 0\ \ 1\ \ 0\ \ 1]$ (即加密後的值)。

　　解密的計算為第 3 層隱藏層到輸出層。其中，第 3 層隱藏層和第 4 層隱藏層之間的權重矩陣為 7 x 7 大小的矩陣 $P^{-1}$ (對應置換矩陣 $P$ 的反矩陣)，計算如公式(18)所示；第 4 層隱藏層和第 5 層隱藏層之間的權重矩陣為 7 x 4 大小的矩陣 $R^{-1}$ (對應解碼矩陣 $R$)，計算如公式(19)所示；第 5 層隱藏層和輸出層之間的權重矩陣為 4 x 4 大小的矩陣 $S^{-1}$ (對應擾碼器矩陣 $S$ 的反矩陣)，計算如公式(20)所示。當第 3 層隱藏層 $y$ 的值為 $[0\ \ 1\ \ 1\ \ 0\ \ 1\ \ 0\ \ 1]$，輸出層的值為$[1\ \ 0\ \ 0\ \ 0]$ (即解密後的值)。

### 三、後量子密碼學神經網路的核心設計

　　在第參.二節中已經表示 McEliece 密碼學方法可以對應到神經網路，所以本研究在此基礎上改進 McEliece 密碼學方法。在每一層的神經網路可以採用不同的非線性激活函數來計算，並且在密文層(如圖 3 中的第 3 層隱藏層)可以加入隨機數擾動，讓密文分佈可以更趨近於均勻分佈，提升安全性。在密文層到輸出層之間，通過神經網路的計算可以去除隨機數擾動的影響，並且還原回明文結果，即完成解密。

　　以第參.二節所提的例子說明，可以建立後量子密碼學神經網路的神經網路結構，如圖 3 所示。為對應上述案例，加密的神經網路結構包含輸入層為 4 個神經元(即 $X = \{x_1, x_2, …, x_c\}$，神經元數量 $c = 4$)、第 1 層隱藏層為 4 個神經元、第 2 層隱藏層為 7 個神經元、以及第 3 層隱藏層為 7 個神經元。其中，輸入層和第 1 層隱藏層之間的權重矩陣為矩陣 $S$、第 1 層隱藏層和第 2 層隱藏層之間的權重矩陣為矩陣 $G$、第 2 層隱藏層和第 3

層隱藏層之間的權重矩陣為矩陣 $P$，並且可以通過非線性激活函數(以⊗表示)計算出密文層(即第 3 層隱藏層)的 $n$ 個神經元值(即 $Y = \{y_1, y_2, …, y_n\}$)，如公式(21)所示。為提升安全性，本研究在密文層輸出值加上隨機數進行擾動；例如，第 3 層隱藏層第 $i$ 個神經元的值為 $y_i$，並且可以加入均勻分佈的隨機數 $r_i$，搭配權重$\alpha$操作隨機數的影響力，可對密文造成擾動，擾動後密文為 $Y' = \{y_1', y_2', …, y_n'\}$，如公式(22)所示。解密的神經網路結構包含第 4 層隱藏層為 7 個神經元、第 5 層隱藏層為 4 個神經元、以及輸出層為 4 個神經元。其中，第 3 層隱藏層和第 4 層隱藏層之間的權重矩陣為矩陣 $L$、第 4 層隱藏層和第 5 層隱藏層之間的權重矩陣為矩陣 $M$、第 5 層隱藏層和輸出層之間的權重矩陣為矩陣 $N$，並且可以通過非線性激活函數計算出輸出層的 $c$ 個神經元值(即 $O = \{o_1, o_2, …, o_c\}$)，如公式(23)所示。除此之外，預期輸出層神經元的值 $O$ 盡可能等於對應的輸入層神經元的值 $X$，並以平均平方誤差(Mean-Square Error, MSE)作為損失函數的一部分進行最佳化。

$$Y = X \otimes S \otimes G \otimes P \tag{21}$$
$$y_i' = y_i + \alpha \times r_i \tag{22}$$
$$O = Y' \otimes L \otimes M \otimes N \tag{23}$$

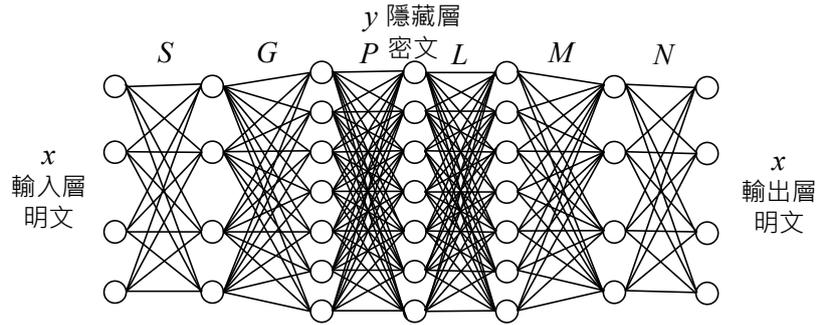

圖 3、後量子密碼學神經網路

　　為提供足夠混亂的密文，本研究分析密文層(如圖 3 中的第 3 層隱藏層)的值是否服從均勻分佈。假設密文層總共有 $n$ 個神經元，對加入隨機數 $r_i$ 後神經元的值 $y_i'$ 做正規化為 0~1 之間為 $h_i$。之後劃分為 $m$ 個區間($m < n$)，計算對每個正規化的值所屬的區間，再統計每個區間的機率，如：第 $j$ 個區間機率值為 $p_j$。本研究採用卡方分佈的累加分配函數值(即 $1 - p$-value 值)分析密文層的輸出值與完全均勻分佈比對，用以判斷是否服從均勻分佈。其中，卡方值$\chi^2$和自由度為 $m - 1$ 的卡方分佈累加分配函數值計算方式如公式(24)和公式(25)所示，並且$\Gamma\left(\frac{m-1}{2}\right)$為 Gamma 函數、$\gamma\left(\frac{m-1}{2}, \frac{\chi^2}{2}\right)$為不完全 Gamma 函數(incomplete gamma function)。當累加分配函數值越小且小於 0.05 則表示越趨近於均勻分佈。因此，本研究將卡方分佈累加分配函數值作為損失函數的一部分，如公式(26)所示。

$$\chi^2 = \sum_{j=1}^{m} \frac{\left(p_j - \frac{1}{m}\right)^2}{\frac{1}{m}} = m \sum_{j=1}^{m} \left(p_j - \frac{1}{m}\right)^2 \tag{24}$$

$$\theta = \frac{\gamma\left(\frac{m-1}{2}, \frac{\chi^2}{2}\right)}{\Gamma\left(\frac{m-1}{2}\right)} \tag{25}$$

$$F(X) = \theta + \frac{1}{c}\sum_{i=1}^{c}(o_i - x_i)^2 \tag{26}$$

本研究提出的後量子密碼學神經網路,可以採用矩陣集合{S, G, P}作為公鑰加密使用、矩陣集合{L, M, N}作為私鑰解密使用,通過神經網路計算,產出的密文可以具有隨機數的擾動,並且密文可以服從均勻分佈,提供安全性的密文。除此之外,可以通過解密的神經網路結構去除隨機數的擾動,並且還原為原本的明文。在此節中雖然只以較小的神經網路結構示意,但可擴展為深度更深、廣度更廣的億級神經元數量來加密更多的資訊和提供更高的安全性。

## 肆、實驗結果與討論

為驗證本研究提出的後量子密碼學神經網路,將採用文獻(Wu, Chen & Zhang, 2019)的資料進行實證分析。該文獻中所使用的資料為行動路訊號資料,使用本研究提出的後量子密碼學神經網路,可以加密行動路訊號。其中,輸入層和輸出層的神經元數分別為361個(即361個基地台),並且設定密文層神經元數量64個,即加密後為64個數字。

由於為讓產出的密文可以足夠混亂,加入隨機數 $r_i$,並且權重$\alpha$為隨機數的權重。本節將從不同的權重值來探討對損失函數中的平均平方誤差和卡方分佈累加分配函數值的影響,如表1所示。其中,為進行比較實驗,所以採用相同的亂數種子產生隨機數,所以每個實驗中隨機數分佈的CDF值都是0.009227 < 0.05,服從均勻分佈。從實驗結果可以發現,當權重$\alpha$越小,則代表隨機數影響越小,所以產生出來的密文較不服從均勻分佈;例如,權重$\alpha = 0.1$時,Y'值分佈的CDF值為0.303414 > 0.5,不服從均勻分佈,但有較低的平均平方誤差(也就是還原明文時誤差較小)。當權重$\alpha \geq 0.4$時,Y'值分佈的CDF值為0.038992 < 0.5,服從均勻分佈;並且隨著權重$\alpha$增加,平均平方誤差將會增加。因此,在本例中,建議採用$\alpha = 0.4$,可以讓密文服從均勻分佈,在還原明文時可以有較低的誤差。

## 伍、結論與未來研究

本研究提出後量子密碼學神經網路,在後量子密碼學 McEliece 密碼學基礎上建構神經網路,並且結合非線性激活函數、隨機數擾動密文、以及約束密文呈均勻分佈等來提升密文的安全性。在實驗中採用行動網路訊號進行實證,並且證明本研究提出的後量子密碼學神經網路可以提供服從均勻分佈的密文,並維持在低平均平方誤差。雖然在第參節方法說明和第肆節實驗證明僅採用較小型的神經網路(輸入層有數百個神經元),但本研究方法可以擴展神經網路規模,可以提升至萬級或億級的神經元數量來進行加解密計算。

未來研究可以考慮設計更有效率且更安全的激活函數來減少加解密的計算時間和密文的安全性。並且可以考慮把本研究提出的後量子密碼學神經網路應用到各個領域。

表 1、後量子密碼學神經網路

| 權重$\alpha$ | 輸出層的平均平方誤差 | 檢定隨機數分佈的 CDF 值 | 檢定 $Y'$ 值分佈的 CDF 值 |
|---|---|---|---|
| 0.1 | 3.74E-05 | 0.009227 | 0.303414 |
| 0.2 | 5.86E-05 | 0.009227 | 0.202276 |
| 0.3 | 6.84E-05 | 0.009227 | 0.089382 |
| 0.4 | 8.27E-05 | 0.009227 | **0.038992** |
| 0.5 | 0.0001 | 0.009227 | **0.018034** |
| 0.6 | 0.0001 | 0.009227 | **0.013643** |
| 0.7 | 0.0002 | 0.009227 | **0.016572** |
| 0.8 | 0.0002 | 0.009227 | **0.012751** |
| 0.9 | 0.0003 | 0.009227 | **0.017519** |
| 1 | 0.0003 | 0.009227 | **0.017685** |

## 參考文獻